\documentclass[sigconf]{acmart} 
\usepackage{comment}
\usepackage{algorithm}
\usepackage[noend]{algpseudocode}

\AtBeginDocument{%
  \providecommand\BibTeX{{%
    \normalfont B\kern-0.5em{\scshape i\kern-0.25em b}\kern-0.8em\TeX}}}

\copyrightyear{2020}
\acmYear{2020}
\setcopyright{rightsretained}
\acmConference[ICAIF '20]{ACM International Conference on AI in Finance}{October 15--16, 2020}{New York, NY, USA}
\acmBooktitle{ACM International Conference on AI in Finance (ICAIF '20), October 15--16, 2020, New York, NY, USA}
\acmDOI{10.1145/3383455.3422570}
\acmISBN{978-1-4503-7584-9/20/10}

\begin{document}

\title{Multi-Agent Reinforcement Learning in a Realistic Limit Order Book Market Simulation}


\author{Micha\"el Karpe}
\authornote{Both authors contributed equally to this research.}
\email{michael.karpe@berkeley.edu}
\author{Jin Fang}
\authornotemark[1]
\email{jin\_fang@berkeley.edu}
\affiliation{
  \institution{University of California, Berkeley}
  \city{Berkeley}
  \state{California}
}

\author{Zhongyao Ma}
\authornote{Both authors contributed equally to this research.}
\email{mazy@berkeley.edu}
\author{Chen Wang}
\authornotemark[2]
\email{chenwang@berkeley.edu}
\affiliation{
  \institution{University of California, Berkeley}
  \city{Berkeley}
  \state{California}
}

\renewcommand{\shortauthors}{Karpe and Fang, et al.}

\begin{abstract}

Optimal order execution is widely studied by industry practitioners and academic researchers because it determines the profitability of investment decisions and high-level trading strategies, particularly those involving large volumes of orders. However, complex and unknown market dynamics pose significant challenges for the development and validation of optimal execution strategies. In this paper, we propose a model-free approach by training Reinforcement Learning (RL) agents in a realistic market simulation environment with multiple agents. First, we configure a multi-agent historical order book simulation environment for execution tasks built on an Agent-Based Interactive Discrete Event Simulation (ABIDES) \cite{byrd2019abides}. Second, we formulate the problem of optimal execution in an RL setting where an intelligent agent can make order execution and placement decisions based on market microstructure trading signals in High Frequency Trading (HFT). Third, we develop and train an RL execution agent using the Double Deep Q-Learning (DDQL) algorithm in the ABIDES environment. In some scenarios, our RL agent converges towards a Time-Weighted Average Price (TWAP) strategy. Finally, we evaluate the simulation with our RL agent by comparing it with a market replay simulation using real market Limit Order Book (LOB) data.
\end{abstract}


\maketitle

\section{Introduction}

\subsection{Agent-based Simulation for Reinforcement Learning in High-Frequency Trading}

Simulation techniques lay the foundations for understanding market dynamics and evaluating trading strategies for both financial sector investment institutions and academic researchers. Current simulation methods are based on sound assumptions about the statistical properties of the market environment and the impact of transactions on the prices of financial instruments. Unfortunately, market characteristics are complex and existing simulation methods cannot replicate a realistic historical trading environment. The trading strategies tested by these simulations generally show lower profitability when implemented in real markets. It is therefore necessary to develop interactive agent-based simulations that allow trading strategy activities to interact with historical events in an environment close to reality.

High Frequency Trading (HFT) is a trading method that allows large volumes of trades to be executed in nanoseconds. Execution strategies aim to execute a large volume of orders with minimal adverse market price impact. They are particularly important in HFT to reduce transaction costs. A common practice of execution strategies is to split a large order into several child orders and place them over a predefined period of time. However, developing an optimal execution strategy is difficult given the complexity of both the HFT environment and the interactions between market participants.

The availability of NASDAQ's high-frequency LOB data allows researchers to develop model-free execution strategies based on RL through LOB simulation. These model-free approaches do not make assumptions or model market responses, but rely instead on realistic market simulations to train an RL agent to accumulate experience and generate optimal strategies. However, no existing research has implemented RL agents in realistic simulations, which makes the generated strategies suboptimal and not robust in real markets.

\subsection{Related work}

The use of RL for developing trading strategies based on large scale experiments was firstly studied by Nevmyvaka et al. in 2006 \cite{nevmyvaka2006reinforcement}. The topic has gained popularity in recent years due to the improvements in computing resources, the advancement of algorithms, and the increasing availability of data. HFT makes necessary the use of RL automate and accelerates order placement. Many papers present such RL approaches, such as temporal-difference RL \cite{spooner2018market} and risk-sensitive RL \cite{mani2019applications}.

Although RL strategies have proven their effectiveness, they suffer from a lack of explainability. Thus, the need for explaining these strategies in a business context has led to the development of representations of risk-sensitive RL strategies in the form of compact decision trees \cite{vyetrenko2019risk}. Advances in the development of RL agents for trading and order placement then showed the need to learn strategies in an environment close to a real market one. Indeed, traditional RL approaches suffer from two main shortcomings. 

First, each financial market agent adapts its strategy to the strategies of other agents, in addition to the market environment. This has led researchers to consider the use of Multi-Agent Reinforcement Learning (MARL) for learning trading and order placement strategies \cite{patel2018optimizing} \cite{balch2019evaluate}. Second, the market environment simulated in classical RL approaches was too simplistic. The creation of a standardized market simulation environment for artificial intelligence agent research was then undertaken to allow agents to learn in conditions closer to reality, through the creation of ABIDES \cite{byrd2019abides}. Research works on the metrics to be considered to evaluate the agents of RL in this environment was also supported within the framework of LOB simulation \cite{vyetrenko2019get}. 

As MARL and ABIDES allow a simulation much closer to real market conditions, additional research was conducted to address the curse of dimensionality, as millions of agents compete in traditional market environments. The use of mean field MARL allows faster learning of strategies by approximating the behavior of each agent by the average behavior of its neighbors \cite{yang2018mean}. 

The notion of fairness applied to MARL \cite{jiang2019learning} also brings both efficiency and stability by avoiding situations where agents could act disproportionately in the market, for example by executing large orders. The integration of fairness into MARL \cite{bao2019fairness} has been studied as an evolution of traditional MARL strategies used for example for liquidation strategies \cite{bao2019multi}.

\subsection{Contributions}

Our main contributions in HFT simulation and RL for optimal execution are the following:
\begin{itemize}
    \item We set up a multi-agent LOB simulation environment for the training of RL execution agents within ABIDES. Based on existing functionality in ABIDES, such as the simulation kernel and non-intelligent background agents, we develop several RL execution agents and make adjustments to the kernel and framework parameters to suit the change.
    \item We formulate the problem of optimal execution within an RL framework, consisting of a combination of action spaces, private states, market states and reward functions. To the best of our knowledge, this is the first formulation with optimal execution and optimal placement combined in the action space.
    \item We develop an RL execution agents using the Double Deep Q-Learning (DDQL) algorithm in the ABIDES environment.
    \item We train an RL agent in a multi-agent LOB simulation environment. In some situations, our RL agent converges to the TWAP strategy.
    \item We evaluate the multi-agent simulation with an RL agent trained on real market LOB data. The observed order flow model is consistent with LOB stylized facts.
\end{itemize}

\section{Optimal execution using Double Deep Q-Learning}

\subsection{Optimal execution formulation}

In our work, we allow RL agents not only to choose the order volume to be placed, but also to choose between a market order and one or more limit orders at different levels of the order book. In this section, we describe the states, actions, and rewards of our optimal execution problem formulation.

We define the trading simulation as a $T$-period problem, which is denoted by the times $T_0 < T_1 < \dots < T_N$ with $T_0 = 0$. We focus on the time horizon from 10:00am to 3:30pm for each trading day, in order to avoid the most volatile periods of the trading day. Indeed, we observe that it is difficult to train agents to capture complex market behaviors given limited data. The time interval within each period is $\Delta T = 30$ seconds, so that there is a total of 660 periods within the time horizon we define in a trading day, lasting 5 hours (i.e. $T_N = T / \Delta T = 660$). $P$ represents the price, while $Q$ represents the quantity volume at a certain price in the limit order book. Our optimal execution problem is then formulated as follows:

\begin{enumerate}
    \item \textbf{State $s$}: the state space includes the information on the LOB at the beginning of each period. For each time period, we use a tuple containing the following characteristics to represent the current state:
    \begin{itemize}
        \item \textit{time remaining} $t$: the time remaining after the time period $T_k$. Since we assume that a trade can only take place at the beginning of each period, this variable also contains the number of remaining trading times. The variable is normalized to be in the $[-1, 1]$ range as follows:
        $$t=2 \times \frac{T-t}{T} - 1$$
        \item \textit{quantity remaining} $n$: the quantity of remaining inventory at the time period $T_k$, depending on the initial inventory $N$, which is also normalized:
        $$n=2 \times \frac{N - \sum_{i = 0}^{t}{n_{i}}}{N} - 1$$
    \end{itemize}
    The above state variables are linked to specific execution tasks, called private states. In addition, we also use the following market state variables to capture the market situation at a given point in time:
    \begin{itemize}
        \item \textit{bid-ask spread} $s$: the difference between the highest bid price and the lowest ask price, which is intended to provide information on the liquidity of the asset in the market: $$s=P_{best\_ask} - P_{best\_bid}$$
        \item \textit{volume imbalance} $v_{\text{imb}}$: the difference between the existing order volume on the best bid and best ask price levels. This feature contains information on the current liquidity difference on both sides of the order book, indirectly reflecting the price trend.
        $$v_{\text{imb}} = \frac{Q_{best\_ask} - Q_{best\_bid}}{Q_{best\_ask} + Q_{best\_bid}}$$
        \item \textit{one-period price return}: the log-return of the stock price over two consecutive days measures the short-term price trend. We intend to allow the RL agent to learn short term return characteristics of the stock price.
        $$r_{1} = \log\left(\frac{P_{t}}{P_{t-1}}\right)$$
        \item \textit{t-period price return}: the log-return of the stock price since the beginning of the simulation measures the deviation between the stock price at time $t$ and the initial price at time $0$.
        $$r_{t} = \log\left(\frac{P_{t}}{P_{0}}\right)$$
    \end{itemize}

    \item \textbf{Action $a$:} the action space defines a possible executed order in a given state, i.e. the possible quantity of remaining inventory affected by the order. In this case, the order can be either a market order or a limit order, either from the bid side or from the ask side. Therefore, the action for each state is a combination of the quantity to be executed with the direction for placement. We use the execution choice to indicate the former and the placement choice to represent the latter.
    \begin{itemize}
    \item \textit{Execution Choices:}
    At the beginning of each period, the agent decides on an execution quantity $N_t$ derived from the order quantity $N_{TWAP}$ placed using the TWAP strategy. $a$ is taken in $\{0.1, 0.5, 1.0, \dots, 2.5\}$ and is a scalar the agent chooses from a set of numbers to increase or decrease $N_{TWAP}$:
    $$N_t = a \cdot N_{TWAP}$$
    \item \textit{Placement Choices:}
    The agent chooses one of the following order placement methods:
        \begin{itemize}
            \item choice 0: Market Order
            \item choice 1: Limit Order - place 100\% on top-level of LOB
            \item choice 2: Limit Order - place 50\% on each of top 2 levels 
            \item choice 3: Limit Order - place 33\% on each of top 3 levels
        \end{itemize}
    \end{itemize}
    \item \textbf{Reward $r$:} the reward intends to reflect the feedback from the environment after agents have taken a given action in a given state. It is usually represented by a reward function consisting of useful information obtained from the state or the environment. In our formulation, the reward function $R_{t}$ measures the execution price slippage and quantity.
    $$R_{t} = \left(1 - \frac{|P_{fill} - P_{arrival}|}{P_{arrival}}\right) \cdot \lambda\frac{N_{t}}{N}$$
    where $\lambda$ is a constant for scaling the effect of the quantity component.
\end{enumerate}

\subsection{Double Deep Q-Learning algorithm}

Aiming to achieve the optimal execution policy, RL enables agents to learn the best action to take through interaction with the environment. The agent follows the strategy that can maximize the expectation of cumulative reward. In Q-Learning, it is represented by the function proposed by Mnih et al. \cite{mnih2013playing}:
$$Q(s,a)=\mathbb{E}\left[\sum_{t=0}^{\infty}\gamma^{t} \times R(s,a_t,s')\right]$$

However, the above approach would be redundant with larger dimensions of the state space where states cannot be visited in depth. Thus, instead of directly using a matrix as the Q-function, we can learn a feature-based value function $Q(s, a | \theta)$, where $\theta$ is the weighting factor that is updated by a stochastic gradient descent.

The Q-value with parametric representation can be estimated by multiple flexible means, in which the Deep Q-Learning (DQL) best fits our problem. In the DQL, the Q-function $Q(s, a | \theta)$ is combined with a Deep Q-Network (DQN), and $\theta$ is the network parameters.

The network memory database contains samples with tuples of information recording the current state, action, reward and next state $(s, a, r, s')$. For each period, we generate samples according to an $\varepsilon$-greedy policy and store them in memory. We replace the samples when the memory database is full, following the First-In-First-Out (FIFO) principle, which means that old samples are removed first.

At each iteration, we batch a given quantity of samples from the memory database, and compute the target Q-value $y$ for each sample, which is defined by Mnih et al. \cite{mnih2013playing} as:
$$y=R(s,a)+\gamma \cdot \max_{a}Q(s,a|\theta)$$
The network parameter $\theta$ is updated by minimizing the loss between the target Q-value and the estimated Q-value calculated from the network, based on current parameters.

However, the DQL algorithm suffers from both instability and overestimation problems, since the neural network both generates the current target Q-value and updates the parameters. A common method to solve this issue is to introduce another neural network with the same structure and calculate the Q-value separately, which is called Double Deep Q-Learning (DDQL) \cite{van2016deep}.

In DDQL, we distinguish two neural networks, the evaluation network and the target network, in order to generate the appropriate Q-value. The evaluation network selects the best action $a^*$ for each state in the sample, while the target network estimates the target Q-value. We update the evaluation network parameters $\theta_E$ every period, and replace the target network parameters $\theta_T$ with $\theta_T=\theta_E$ after several iterations.

\section{Reinforcement Learning in ABIDES}

\subsection{ABIDES environment}

ABIDES is an Agent-Based Interactive Discrete Event Simulation environment primarily developed for Artificial Intelligence (AI) research in financial market simulations \cite{byrd2019abides}.

The first version of ABIDES \textit{(0.1)} was released in April 2019 \cite{byrd2019abides}. The ABIDES development team released a second version \textit{(1.0)} in September 2019, which is supposed to be the first stable release. Finally, the latest version \textit{(1.1)}, released in March 2020, adds many new functionalities, including the implementation of new agents, such as Q-Learning agents, as well as the computation of the realism metrics.

ABIDES aims to replicate a realistic financial market environment by largely replicating the characteristics of real financial markets such as NASDAQ, including \textit{nanosecond time resolution, network latency and agent computation delays and communication solely by means of standardized message protocols} \cite{byrd2019abides}. In addition, by providing ABIDES with historical LOB data, we are able to reproduce a given period of this history using ABIDES \textit{marketreplay} configuration file.

ABIDES also aims to help researchers in answering questions raised about the understanding of market behavior, such as the influence of delays in sending orders to an exchange, the price impact of placing large orders or the implementation of AI agents into real markets \cite{byrd2019abides}.

ABIDES uses a hierarchical structure in order to ease the development of complex agents such as AI agents. Indeed, thanks to \textit{Python} \textit{object-oriented programming} and \textit{inheritance}, we can, for example, create a new \textit{ComplexAgent} class which inherits from the \textit{Agent} and thus benefits from all functionalities available in the \textit{Agent} class. We can then use \textit{overriding} if we want to change an \textit{Agent} function in order to make it specific for our \textit{ComplexAgent}.

Given that ABIDES is not only for financial market simulations, the base \textit{Agent} class has nothing related to financial markets and is provided only with functions for Discrete Event Simulation. The \textit{FinancialAgent} class inherits from \textit{Agent} and has supplementary functionalities to deal with currencies. On the one hand, the \textit{ExchangeAgent} class inherits from \textit{FinancialAgent} and simulates an financial exchange. On the other hand, the \textit{TradingAgent} also inherits from \textit{FinancialAgent} and is the base class for all trading agents which will communicate with the \textit{ExchangeAgent} during financial market simulations.

Some trading agents -- i.e. inheriting from the \textit{TradingAgent} class -- are already provided in ABIDES, such as the \textit{MomentumAgent} which places orders depending on a given number of previous observations on the simulated stock price. In the next sections, unless otherwise mentioned, the agents we refer to are all trading agents.

We present in Figure 1 an example of market simulation in ABIDES using real historical data. A \textit{ZeroIntelligenceAgent} places orders on a stock in a market simulation with 100 \textit{ZeroIntelligenceAgent} trading against an \textit{ExchangeAgent}. Each agent is able to place long, short or exit orders, competing with thousands of other agents to maximize their reward.

\begin{figure}[ht]
\centering
\includegraphics[width=\linewidth]{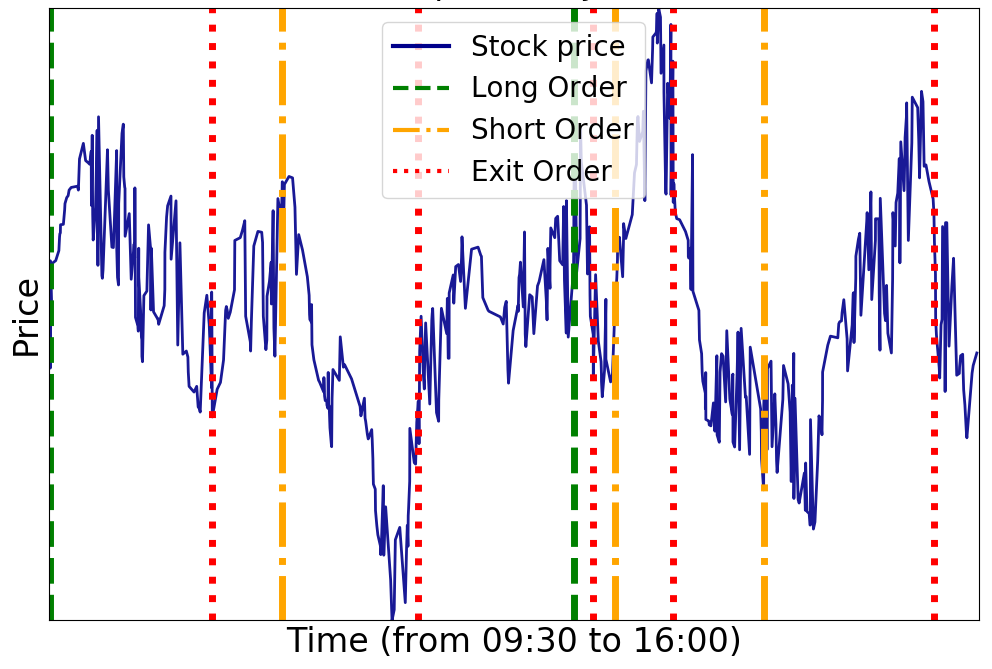}
\caption{Agent placing orders on simulated stock price}
\end{figure}

\subsection{DDQL implementation in ABIDES}

In order to train the DDQL agent in ABIDES during a \textit{marketreplay} simulation, the learning process needs to be integrated with the simulation process. The training process starts by initializing the ABIDES execution simulation kernel and instantiating a \textit{DDQLExecutionAgent} object. The same agent object needs to complete $B$ simulations, $B$ being referred to as the number of training episodes.

Within each training episode, the simulation is divided into $N$ discrete periods. For each period $T_{i}$, the agent chooses an action $a_{T_{i}}$ for the current period according to the $\varepsilon$-greedy policy in order to achieve a balance of exploration and exploitation. Then, an order schedule is generated, based on the quantity and placement strategy defined in the chosen action. The current-period order could be broken into small orders and placed on different levels of the LOB. Then, the current experience $\left(s_{T_{i}}, a_{T_{i}},s_{T_{i+1}}, r_{T_{i}} \right)$ is stored in the replay buffer $\mathcal{D}$.

The replay buffer removes the oldest experience when its size reaches to the maximum capacity specified. This intends to use relatively recent experiences to train the agent. As long as the size of the replay buffer $\mathcal{D}$ reaches a minimum training size, a random minibatch $\left(s_{(j)}, x_{(j)}, r_{(j)}, s_{(j)}^{s, x}\right)$ is sampled from $\mathcal{D}$ for training the evaluation network.

The target network is updated after training the evaluation network 5 times. This training frequency is chosen with a small grid search on multiple training frequencies, balancing computing burden, model fitting performance, and experience accumulation. The final step within time period $T_{i}$ is to update the state $s_{T_{i+1}}$ and compute the reward $r_{T_{i}}$ for the current period. The entire process is summarized in the algorithm below.

\begin{algorithm}
\caption{Training of DDQL for optimal execution in ABIDES.}
\label{euclid}
\begin{algorithmic}[1]
\For{training episode $b \in B$}
	\For{$i \gets 0$ to $N-1$}
		\State With probability $\varepsilon$ select random action $a_{T_{i}}$
		\State Otherwise select optimal action $a_{T_{i'}}$ based on target\_net
		\State Schedule orders $o_{i}$ according to $a_{i}$ and submit $o_{i}$
		\State Store experience $\left(s_{T_{i}}, a_{T_{i}},s_{T_{i+1}}, r_{T_{i}} \right)$ in replay buffer $\mathcal{D}$
		\If {$length(\mathcal{D}) > max\_experience$}
			\State Remove oldest experience
		\EndIf
		\If {$length(\mathcal{D}) \geq min\_experience$ AND $i \mod 5 == 0$}
			\State Sample random minibatch from $\mathcal{D}$ 
			\State Train eval\_net and update target\_net
		\EndIf
		\If {orders $o_{i}$ accepted or executed}
			\State Observe environment and update $s_{T_{i+1}}$
			\State Compute and update $r_{T_{i}}$
		\EndIf
	\EndFor
\EndFor
\end{algorithmic}
\end{algorithm}	

To train a DDQL agent, we implement our neural network based on a Multi Layer Perceptron (MLP). We stack multiple dense layers together as illustrated in Figure 2, and we set the activation function to be ReLU to introduce non-linearity. Dropout is added to avoid overfitting. The optimization algorithm chosen for backpropagation is the \textit{Root Mean Square back-propagation} (RMSprop) proposed by Tieleman et al. \cite{tieleman2012lecture} with a learning rate of 0.01. The loss function we choose is the \textit{mean squared error} (MSE). The size of the output layer size is the number of actions to choose from.

\begin{figure}[ht]
\centering
\includegraphics[width=0.9\linewidth]{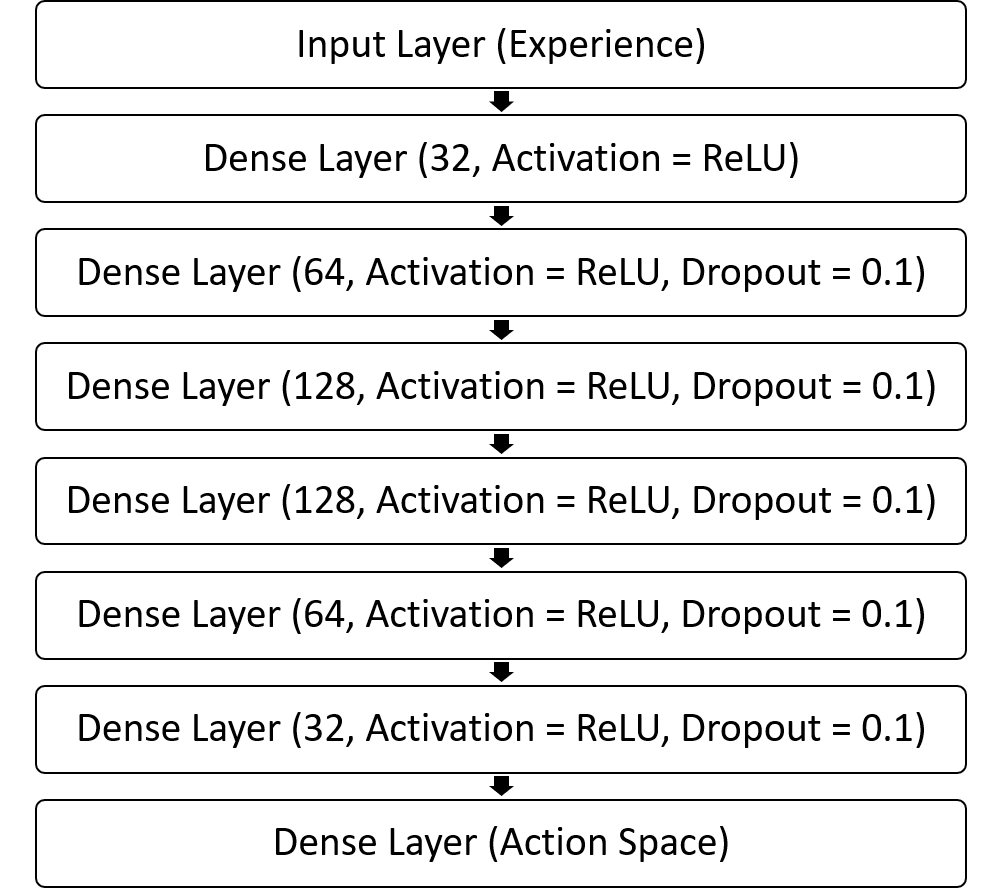}
\caption{Multi-Layer Perceptron (MLP) architecture}
\end{figure}

\section{Experiments and Results}
In this section, we describe our experiment for training a DDQL agent in a multi-agent environment and observe the behavior of the agent during testing.

\subsection{Data for experiments}

The data we use for the experiments are NASDAQ data converted to LOBSTER \cite{huang2011lobster} format to fit the simulation environment. We extract the order flow for 5 stocks (CSCO, IBM, INTC, MSFT and YHOO) from January 13 to February 6, 2003. We train the model over 9 days and test it over the following 9 days. The training data is concatenated into a single sequence, and the training process is continuous for consecutive days while the model parameters are stored in intermediate files.

\subsection{Multi-agent configuration}

The multi-agent environment that we have set up for the training of DDQL agents at ABIDES consists of an \textit{ExchangeAgent}, a \textit{MarketreplayAgent}, six \textit{MomentumAgents}, a \textit{TWAPExecutionAgent} and our \textit{DDQLExecutionAgent}. 

\begin{itemize}
    \item \textit{ExchangeAgent} acts as a centralized exchange that keeps the order book and matches orders on the bid and ask sides. 
    \item \textit{MarketreplayAgent} accurately replays all market and limit orders recorded in the historical LOB data.
    \item \textit{MomentumAgent} compares the last 20 mid-price observations with the last 50 mid-price observations and places a buy limit order if the 20 mid-price average is higher than the 50 mid-price average, or a sell limit order if it is.
    \item \textit{TWAPAgent} adopts the TWAP strategy. This strategy minimizes the price impact by dividing a large order equally into several smaller orders. Its execution price is the average price of the recent $k$ time periods. The agent's optimal trading rate is calculated by dividing the total size of the order by the total execution time, which means that the trading quantity is constant. When the stock price follows a Brownian motion and the price impact is assumed to be constant, this strategy is optimal \cite{daberius2019deep}. In RL, if there is no penalty for any running inventory, but a significant penalty for the ending inventory, the TWAP strategy is also optimal.
\end{itemize}

\subsection{Result of our experiments}

We observe that our RL agent converges to the TWAP strategy after 9 consecutive days of training regardless of the stock chosen. The agent places a top-level limit order or market order. However, the execution quantity chosen by the agent throughout the test period changes with the stock. This result may be explained by the fact that our RL agent places an order every 30 seconds, and thus is not able to capture the trading signals existing during shorter periods of time.

\section{Realism of our LOB simulation}

Numerous research papers studied the behaviour of the LOB. A recent research paper presents a review of these LOB characteristics which can be referred to as \textit{stylized facts}, as reminded in Vyetrenko et al. \cite{vyetrenko2019get} In this section, we compare our simulation with real markets based on these \textit{realism metrics} in order to assess whether our market simulation, mainly built on ABIDES, is realistic.

We can mainly distinguish two sets of metrics for the analysis of the LOB behavior. The first set includes metrics related to asset return distributions and the second set includes metrics related to volume and order flow distributions \cite{vyetrenko2019get}.

Asset returns metrics generally relate to price return or percentage change. For the LOB, it includes the mid-price trend, which is the average of the best bid price and the best ask price. Volumes and order flow metrics, for their part, relate to the behavior of incoming order flows, including new buy orders, new sell orders, order modifications or order cancellations. Three main stylized facts related to order flows are as follows:
\begin{itemize}
    \item \textbf{Order volume in a fixed time interval:} Order volume in a fixed interval or time window likely follows a positively skewed log-normal distribution or gamma distribution \cite{abergel2016limit}.
    \item \textbf{Order interarrival time:} The time interval of two consecutive limit orders likely follows an exponential distribution \cite{li2018generating} or a Weibull distribution \cite{abergel2016limit}. 
    \item \textbf{Intraday volume patterns:} Limit order volume within a given time interval for each trading day can be approximated by a U-shaped polynomial curve, where the volume is higher at the start and the end of the trading day \cite{bouchaud2018trades}.
\end{itemize}

These realism metrics are implemented in ABIDES. We compute the stylized facts implemented in ABIDES on a \textit{marketreplay} simulation with and without the presence of our \textit{DDQLExecutionAgent}. For all of the stylized facts, we observe that adding a single new agent to a simulation does not significantly alter the result of the computation. This means that evaluating the realism of our simulation with a single \textit{DDQLExecutionAgent} is equivalent to evaluating the realism of the LOB data provided as an input to the simulation.

On our NASDAQ LOB 2003 data, we always observe the two first order flow stylized facts mentioned above, but not the intraday volume patterns. As an example of successfully observed stylized fact, Figure 3 illustrates the stylized fact about order volume in a fixed time interval for the IBM stock on January 13, 2003. We verify that order volume in a fixed time interval follows a gamma distribution. It shows that the most likely order volume value in the 60-second window is very low, while the volume average can be high. 

\begin{figure}[ht]
\centering
\includegraphics[width=\linewidth]{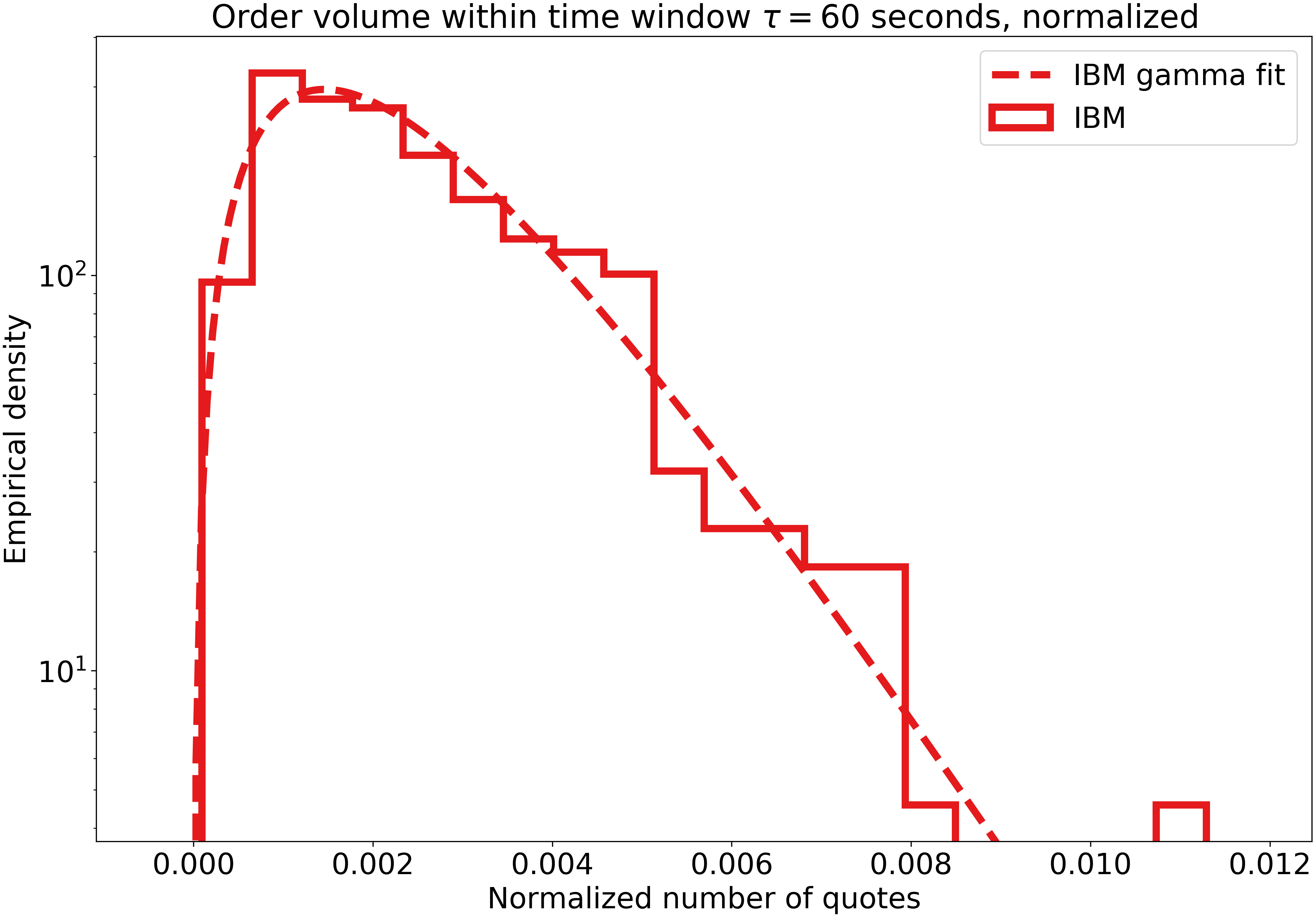}
\caption{Order volume in fixed time interval for IBM stock on January 13, 2003}
\end{figure}

\section{Conclusion and future work}

\subsection{Conclusion}

In this paper, we built our \textit{DDQLExecutionAgent} in ABIDES by implementing our own optimal execution problem formulation through RL in a financial market simulation, and set up a multi-agent simulation environment accordingly. In addition, we conducted experiments to train our \textit{DDQLExecutionAgent} in the ABIDES environment and compared the agent strategy with the TWAP strategy to which our agent converges. Finally, we evaluated the realism of our multi-agent simulations by comparing LOB stylized facts on simulations using our \textit{DDQLExecutionAgent} with the ones of a \textit{marketreplay} simulation using real LOB data.

\subsection{Future work}

Due to limited computing resources and lack of data, the experiments we have been able to conduct are limited. Our current model can be improved in many ways. First, instead of training agents on steady periods (10:00am to 3:30pm) only, we can train agents on entire trading days. The agent period of time that we set can be refined to a time interval closer to the nanosecond, as in real HFT. Second, the action space can be expanded to include more types of execution actions and the reward function can be enhanced to include more information and feedback from both the market and other agents. With respect to the state's feature, we choose log price to explicitly capture the mean reversion in financial market which may easily overestimate the effect. Therefore, a more appropriate indicator should be chosen to modify this situation.

Regarding the RL algorithm, we can try several advanced methods to implement an updated approach on our \textit{DDQLExecutionAgent}. Directions include the use of prioritized experience replay to increase the frequency of batching important transitions from memory, or the combination of bootstrapping with DDQL to improve exploration efficiency. Other methods such as increasing the complexity of the network architecture will only be useful if implemented in a more complex environment. Several LSTM layers can be added to take advantage of the agent's past experience and improve the performance when providing a larger training data set or applying a longer training time period.

So far, we focused on a relatively monotonous set of multiple agents, which is not able to fully capture the influence of the interaction between the agents. To remedy this situation, more and different types of agents can be added to the configuration to study collaboration and competition among agents in further detail. Moreover, after introducing a more complex combination of agents in the ABIDES environment, we can try to perform the financial market simulation on the basis of this configuration, which should be much more realistic than the existing one.

The approach to evaluation is also an aspect that can be further expanded. Our experience does not currently allow us to clearly distinguish the difference between our agents and the benchmark. In order to assess the model more accurately, we can further improve our evaluation methods to examine both the parameters of RL and financial performance. For example, by conducting a simulation of the financial market in the ABIDES environment, we can use the realism metrics we have designed to evaluate our agents.

In our experiments, our \textit{DDQLExecutionAgent} learns how to perform a TWAP strategy because its trading frequency is not high enough. However, this work shows the potential of MARL for developing optimal trading strategies in real financial markets, by implementing agents with an higher trading frequency in realistic market simulations.

\begin{acks}
We would like to thank Prof. Xin Guo and Yusuke Kikuchi for their helpful comments during the realization of this work. We also want to thank Svitlana Vyetrenko, Ph.D. for her answers to our questions about the work she contributed to and cited in the References section.
\end{acks}

\bibliographystyle{ACM-Reference-Format}
\bibliography{ms}

\end{document}